\documentclass[twocolumn]{article}
\usepackage[section]{placeins}

\usepackage{graphicx}
\usepackage{float}
\usepackage{amssymb}
\usepackage{amsmath}
\usepackage{mathtools}
\usepackage{bm}
\usepackage[colorlinks,linkcolor=cyan,citecolor=cyan]{hyperref}
\usepackage{extarrows}
\usepackage{cite}
\usepackage{multirow}
\usepackage{colortbl}
\usepackage{graphicx}
\usepackage{multirow}
\usepackage{longtable}

\definecolor{shadecolor}{rgb}{0.9, 0.9, 0.9}  % Define a light gray color for shading
\graphicspath{{figure/}}
\usepackage[a4paper]{geometry}
\title{Conditional Flood Fill Method in Logic Synthesis}
\author{jun jiang}
\date{February 2023}

\begin{document}
\bibliographystyle{unsrt}
\author{Shitian Yang, Junyue Jiang, Yilai Liang, Xiaoyang Chu }

\maketitle
\begin{abstract}
 In the field of Electronic Design Automation (EDA), logic synthesis plays a pivotal role in optimizing hardware resources. Traditional logic synthesis algorithms, such as the Quine-McCluskey method, face challenges in scalability and efficiency, particularly for higher-dimension problems. This paper introduces a novel heuristic algorithm based on Conditional Flood Fill Method aimed at addressing these limitations. Our method employs count-based adjacent element handling and introduces nine new theorems to guide the logic synthesis process. Experimental results validate the efficacy of our approach, showing significant improvements in computational efficiency and scalability compared to existing algorithms. The algorithm holds potential for future advancements in circuit development and Boolean function optimization.
\end{abstract}

\section{Background}

Electronic Design Automation (EDA) serves as the foundational hardware layer for neural networks and artificial intelligence (AI), and is intrinsically tied to the domain of logic synthesis \cite{testa2020extending, mccarthy2006proposal}. As per Moore's Law, the computational capabilities are growing exponentially, further catalyzed by a data processing rate that is predicted to reach a staggering 175 ZB by 2025 \cite{chen2014data}. This explosive growth necessitates relentless advancements in computational speed and efficiency. As we approach the physical limits of traditional hardware, the spotlight shifts towards algorithmic innovations in the realm of logic optimization.

Historically, two-level logic synthesis, as proposed by Shannon, targeted the minimization of sum-of-products (SOP) expressions \cite{r1}. One of the seminal algorithms in this field is the Quine-McCluskey (Q-M) algorithm, established in 1949. While it has been widely adopted for logic optimization, it comes with inherent limitations, such as scalability issues and lack of optimality guarantees \cite{mccluskey1956minimization, quine1952problem}. Specifically, the Q-M algorithm employs an iterative process of partitioning and merging elements in Karnaugh maps, a method that becomes increasingly inefficient as the dimension approaches twenty \cite{karnaugh1953map}. To address some of these shortcomings, the Espresso heuristic algorithm was introduced in 1982. Although Espresso markedly reduced both the time and storage requirements, it, too, lacks guarantees for finding the optimal solution and calls for further refinements, particularly for higher-level logic minimization \cite{brayton1982newton, theobald1998fast}.

In the context of the Q-M algorithm, the procedure involves three primary steps. First, all minterms are identified and represented as the smallest units in a Karnaugh map. Next, these minterms are continuously merged if they differ by a single literal, until prime implicants are formed. Finally, essential prime implicants are selected to cover all minterms, thereby forming the simplest SOP expression. Although theoretically elegant, the Q-M algorithm suffers from computational inefficiencies, especially when dealing with higher-dimension problems. For instance, during the third step, the algorithm traverses multiple paths indiscriminately to cover all minterms, thereby sacrificing both efficiency and accuracy.

This paper introduces a heuristic algorithm that offers superior efficiency and accuracy compared to the aforementioned algorithms for logic synthesis problems involving fewer than twenty variables. Additionally, we contribute nine original theorems that hold promising implications for future advancements in circuit development.

\section{Introduction}
\subsection{Preliminaries}

We introduce several foundational concepts that underpin the methodologies discussed in this paper.

\begin{itemize}
    \item \textbf{Inverse Representation}: The prime symbol (') is used to denote the inverse of a variable.  For a given Boolean variable \( X \), the inverse state is denoted as \( X' \). This notation is used to represent the complement or negation of the variable.
    \begin{itemize}
        \item \textit{Example}: If \( X \) represents a Boolean state, \( X' \) denotes its inverse. Typically, \( X = 1 \), then \( X' = 0 \).
    \end{itemize}
    \item \textbf{Application in Functions}: In Boolean functions, the prime notation is utilized to easily identify the complemented variables within expressions.
    \begin{itemize}
        \item \textit{Example}: In the function \( f(A,B,C,D) = A'C' + A'BD \), the terms \( A' \) and \( C' \) are the inverses of \( A \) and \( C \), respectively.
    \end{itemize}
\end{itemize}
\begin{itemize}
    \item \textbf{Boolean Coordinate}: In an \( n \)-dimension K-map, each element is assigned a unique Boolean coordinate, an \( n \)-bit string representing the state of each of the \( n \) variables.
    \begin{itemize}
        \item \textit{Example}: Refer to Table.1's TRUTH TABLE for \( f \), where \( ABCD' \) is assigned the Boolean coordinate \( 1110 \).
    \end{itemize}
    \item \textbf{Adjacent Elements}: Elements in a K-map are termed adjacent if their Boolean coordinates differ by exactly one bit.
    \begin{itemize}
        \item \textit{Example}: As shown in Table.2's TRUTH TABLE for \( f \), \( 0000 \) and \( 0001 \) are adjacent elements.
    \end{itemize}
    \item \textbf{Dimensionality in Boolean Functions}: Each variable in a Boolean function adds a new dimension to the function's representation. For instance, \( f(A,B,C,D) = A'C' + A'BD \) is a function in four dimensions.
    \item \textbf{Minterms}: Minterm is  the product form of \(m\) variables, where \(m\) is no greater than the total dimensions of the Boolean function denoted as \(n\). Each variable will occur only once either in its form of \( X \) or its inverse form of \( X' \) within the minterm. 
    \item \textbf{Implicants}: Implicants are minterms in a Boolean function's SOP (Sum of Products) representation.
    \begin{itemize}
        \item \textit{Example}: For \( f(A,B,C,D) = A'C' + A'BD \), the implicants are \( A'C' \) and \( A'BD \).
    \end{itemize}
\end{itemize}
% First table from the first graph
\begin{table}[h]
    \centering
    \begin{tabular}{|c|c|c|c|c|}
        \hline
        A & B & C & D & OUTPUT \\
        \hline
        0 & 0 & 0 & 0 & 1 \\
        \hline
        0 & 0 & 0 & 1 & 1 \\
        \hline
        0 & 0 & 1 & 0 & 0 \\
        \hline
        0 & 0 & 1 & 1 & 0 \\
        \hline
        0 & 1 & 0 & 0 & 1 \\
        \hline
        0 & 1 & 0 & 1 & 1 \\
        \hline
        0 & 1 & 1 & 0 & 0 \\
        \hline
        0 & 1 & 1 & 1 & 1 \\
        \hline
        1 & 0 & 0 & 0 & 0 \\
        \hline
        1 & 0 & 0 & 1 & 0 \\
        \hline
        1 & 0 & 1 & 0 & 0 \\
        \hline
        1 & 0 & 1 & 1 & 0 \\
        \hline
        1 & 1 & 0 & 0 & 0 \\
        \hline
        1 & 1 & 0 & 1 & 0 \\
        \hline
        1 & 1 & 1 & 0 & 0 \\
        \hline
        1 & 1 & 1 & 1 & 0 \\
        \hline
    \end{tabular}
    \caption{Truth table for the boolean function with inputs A, B, C, and D.}
\end{table}

% Second table from the second graph
\begin{table}[h]
    \centering
    \begin{tabular}{|c|c|c|c|c|}
        \hline
        CD/AB & 00 & 01 & 11 & 10 \\
        \hline
        00 & 1 & 1 & 0 & 0 \\
        \hline
        01 & 1 & 1 & 0 & 0 \\
        \hline
        11 & 0 & 1 & 0 & 0 \\
        \hline
        10 & 0 & 0 & 0 & 0 \\
        \hline
    \end{tabular}
    \caption{Truth table based on the inputs AB and CD.}
\end{table}
\subsection{Dimension analysis of Boolean function}
For a specific given Boolean function of dimension \(n\), all possible values in Boolean space can be denoted as \{0, 1\}\(^n\). As depicted in Figure. 1, vertices represent all possibilities of minterms with \(n\) variables. 
\begin{figure}
    \centering
    \includegraphics[width=0.9\linewidth]{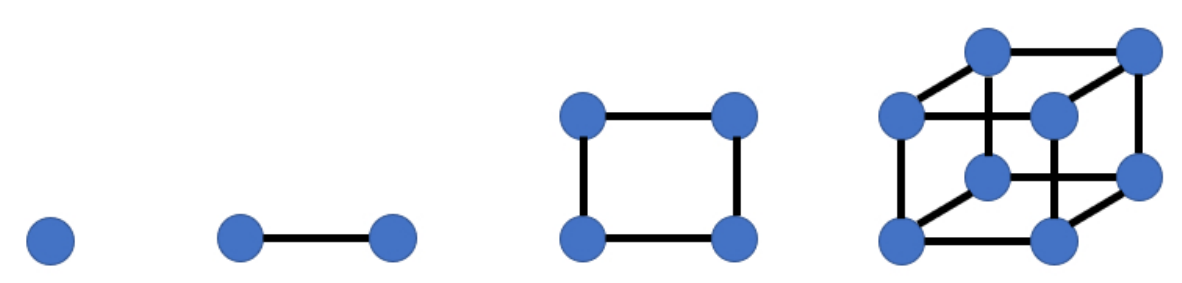}
    \caption{Dimension representation of Boolean space for \(n = 0, 1, 2, 3\)}
    \label{Figure 1}
\end{figure}
Vertices can also be further assigned with dimensional coordinates as shown in Figure. 2, e.g. vertex \((1, 0, 0)\) represents the minterm of \( X\overline{Y}\overline{Z} \). Vertices connected by an edge are adjacent elements differing by exactly one bit. If one Boolean space and its adjacent Boolean space of the same dimension are both minterms, then they can be combined together for simplification, e.g. two adjacent edges can form a square. When a minterm only consist of \(m\) variables, its current Boolean space be of dimension \(n - m\), with a total of \( 2^{n-m} \) vertices.

\begin{figure}
    \centering
    \includegraphics[width=1\linewidth]{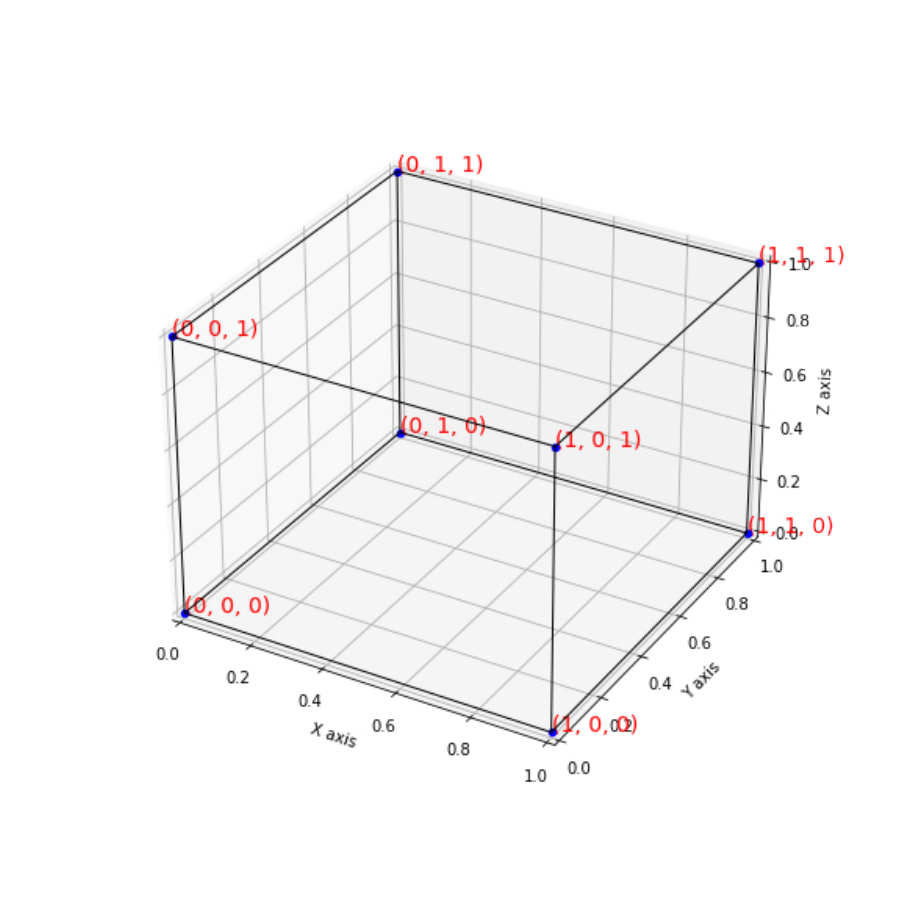}
    \caption{Boolean function in dimension coordinates (\(n = 3)\)}
    \label{Figure 2}
\end{figure}
\subsection{Rules and Deductions}

Our proposed method diverges from the conventional Quine-McCluskey algorithm by employing a count-based approach to handle adjacent elements. Specifically, the first step entails counting the number of adjacent elements for each element and storing this information in a data structure, commonly a dictionary. Subsequently, elements are sorted in descending order based on their adjacent count. To validate this approach, we present the following deductions and proofs:

\paragraph{Rule 1: Coordinate-Boolean Mapping} 
The common bits in the Boolean coordinates of all elements in an implicant correspond to the literals in the representative Boolean expression. For instance, in Fig-3, the shaded portion represents \( A'C' \) in Boolean algebra.

\textit{Proof:} 
Let \( O \) be the input set and \( X, Y \subset O \). From our definition of an implicant, if \( \forall x \in X, x = 1 \) and \( \forall y \in Y, y = 0 \), then \( X\overline{Y} \) is an implicant of \( O \). Given \( F(A,B,C,D) = 1 \), if \( A = 0 \) and \( C = 0 \), then \( \overline{AC} \) is an implicant of \( F \).

\begin{table}[h]
    \centering
    \begin{tabular}{|c|c|c|c|c|}
        \hline
        CD/AB & 00 & 01 & 11 & 10 \\
        \hline
        00 & \cellcolor{shadecolor}1 & \cellcolor{shadecolor}1 & 0 & 0 \\
        \hline
        01 & \cellcolor{shadecolor}1 & \cellcolor{shadecolor}1 & 0 & 0 \\
        \hline
        11 & 0 & 1 & 0 & 0 \\
        \hline
        10 & 0 & 0 & 0 & 0 \\
        \hline
    \end{tabular}
    \caption{Truth table based on the inputs AB and CD with specific cells shaded.}
\end{table}
\paragraph{Rule 2: Cardinality of an Implicant} 
An implicant can contain at most \( 2^A \) elements.

\textit{Proof:} 
Base Case: An implicant differing in one dimension comprises \( 2^1 \) elements.
Induction Step: Assume an \( n \)-dimension implicant contains \( 2^n \) elements. Then, an \( (n+1) \)-dimension implicant will contain \( 2 \times 2^n \) elements.

\paragraph{Rule 3: Coordinate Similarity in N-Dimensions}
In \( N \)-dimensions, if \( 2^A \) elements can be combined, then their \( N-A \) coordinates are identical.

\textit{Proof:} 
This follows directly from Rule 1. These elements will differ in \( A \) coordinates, implying that the remaining \( N-A \) coordinates must be the same.

\paragraph{Rule 4: Covering by Adjacency}
An element with a higher count of adjacent elements can be covered by more implicants.

\paragraph{Rule 5: Adjacency Limit in N-Dimensions}
In \( N \)-dimensions, an element can have at most \( N \) adjacent elements.

\textit{Proof:}
In a binary numbering system, each adjacent element differs in one bit. Hence, an \( N \)-dimension element can have at most \( N \) adjacent elements.

\paragraph{Rule 6: Lower Bound on Adjacency}
Any element in an implicant with \( 2^A \) elements must have at least \( A \) adjacent elements.

\subsection{\textit{Proof:}}
This is a corollary from Rule 3. An element in a \( 2^A \)-element implicant will differ in \( A \) coordinates from its adjacent elements, leading to at least \( A \) adjacent elements.

\paragraph{Rule 7: Upper Bound on Implicant Size}
If an element has \( A \) adjacent elements, the largest implicant containing this element has \( 2^A \) elements.

\textit{Proof:}
Contradiction method. Assuming the existence of an element \( x_0 \) with \( A \) adjacent elements that belongs to an implicant \( I \) with more than \( 2^A \) elements contradicts Rule 6, thus proving the assertion.

\paragraph{Rule 8: Implicant Determination}
An element and its \( A \) adjacent elements can uniquely determine an implicant that can cover them.

\textit{Proof:}
This is directly inferred from Rule 7.

\paragraph{Rule 9: Conditional Combination}
Once an element is covered , it becomes a "don't care" condition, allowing for its optional combination with other elements.

\section{Algorithm Theory}
\subsection{Preliminary Work}
The scope of the proposed algorithm is confined to single-output Boolean functions, represented in Sum of Products (SOP) form. 

\paragraph{Data Initialization}
Initially, the algorithm parses the truth table to extract all outputs manifesting a '1' value. Concurrently, it calculates their Boolean coordinates by converting their positional index in the truth table to binary form. This operation aligns with Rule 1, which articulates the formulation of Boolean coordinates.

\paragraph{Neighbor Extraction}
For each such '1' output, its adjacent elements are determined by varying a single coordinate as definition. The set of all possible neighboring elements for each '1' output is stored in a set denoted as \textbf{NE}.

\paragraph{Neighbor Count and Sorting}
The cardinality of each \textbf{NE} set is then calculated and stored in a corresponding set \textbf{NB}. To minimize the inclusion of unnecessary implicants, the '1' outputs are sorted based on their \textbf{NB} values in ascending order, as per Rule 4, and stored in a primary set named \textbf{MAIN}. This sorting strategy is essential to avoid a situation where processing implicants with a higher \textbf{NB} count first might lead to them being superseded by later processed implicants with a lower \textbf{NB} count, resulting in the introduction of unnecessary implicants.

\subsection{Main Algorithmic Procedure}
The proposed algorithm systematically computes implicants by sequentially processing the coordinates of elements within the \textbf{MAIN} set.

\paragraph{Initial Implicant Identification}
Initially, the algorithm selects a 'main element' from \textbf{MAIN} and identifies \( k \) neighboring elements in a random manner. The selection prioritizes elements not yet encompassed by any implicant. The choice of \( k \) is iterative-dependent, commencing with the \textbf{NB} value of the main element, adhering to Lemma 7. This is aimed at identifying the largest possible implicant. Subsequently, leveraging Lemma 8, a potential implicant expression is derived using these \( k+1 \) elements (inclusive of the main element and its \( k \) neighbors).

\paragraph{Implicant Expansion Process}
The algorithm adopts a flood-fill-like strategy to explore additional elements potentially constituting the implicant. It initializes two sets: \textbf{SURE} and \textbf{CHECK}. Initially, the \( k+1 \) elements are placed in \textbf{SURE}, while their unprocessed neighboring elements are added to \textbf{CHECK}. Each element in \textbf{CHECK} is evaluated for its compatibility with the prospective implicant expression and checked against Lemma 6 to ascertain if it exceeds \( k \) neighbors. Elements meeting these criteria are transferred to \textbf{SURE}, with their unprocessed neighbors shifted to \textbf{CHECK}. Should an element have \( k \) or fewer neighbors but aligns with the implicant, the process is aborted, \( k \) is decremented, and a new iteration commences.

\paragraph{Implicant Validation and Finalization}
In the concluding step, the algorithm scrutinizes the \textbf{SURE} set in accordance with Lemmas 7 and 8 to confirm if its size equals \( 2^k \). A size smaller than \( 2^k \) indicates a failed implicant, prompting a decrement in \( k \) and a subsequent reiteration with a new set of \( k \) neighboring elements. Upon successful validation, the implicant's expression is cataloged in the \textbf{OUTPUT} set. Elements within \textbf{SURE} are then expunged from \textbf{MAIN}, as dictated by Lemma 9.

The algorithm proceeds until all elements in \textbf{MAIN} have been evaluated, culminating in the generation of the \textbf{OUTPUT} set, which encompasses all validated implicants.

\subsection{Algorithmic Soundness Analysis}
\subsubsection{Elimination of Redundant Implicants}

In this methodology, redundant implicants are notably absent. \\
\textbf{Proof}: \\
Suppose, counter to our claim, that a redundant implicant exists, hereafter denoted as \( \text{UI} \). This implicant cannot serve as a sub-implicant to a larger covering implicant. The rationale behind this is straightforward: our primary algorithm is engineered to locate the largest covering implicants for each constituent element.

Let \( \text{UI} \) be covered by implicants in the sets \( \{ A_1, A_2, \ldots, A_q \} \) and \( \{ B_1, B_2, \ldots, B_p \} \), where each \( A_i \) is smaller than \( \text{UI} \) and each \( B_i \) is larger than \( \text{UI} \). According to the sequence of operations in the primary algorithm, \( A_i \) is processed prior to \( \text{UI} \), and \( \text{UI} \) precedes \( B_i \). When \( \text{UI} \) is being computed, a main element—designated \( \text{ER} \)—is selected from within \( \text{UI} \). This \( \text{ER} \) will unavoidably be covered by one of the \( B_i \), thereby generating a contradiction. Hence, the existence of redundant implicants is refuted.

\subsubsection{Validity of Primary Procedure}

\textbf{Completeness}: To ensure full coverage of all elements, each element in the \textbf{MAIN} set is scrutinized. An element is extracted from \textbf{MAIN} only when it has been subsumed in an implicant. \\
\textbf{Conformity}: All the elements within the computed implicants strictly adhere to the condition that neighbors can only have outputs of "1". Consequently, elements with an output of "0" are explicitly excluded from the \textbf{SURE} set during the search algorithm. \\
\textbf{Local Optimality}: Implicants are chosen based on the descending order of their neighbor counts, in accordance with Rules 2 and 7. This ensures the identification of the largest implicant for each individual element. \\
\textbf{Limitations}: The algorithm’s focus on the largest implicant for each element may not always lead to a global optimum. Strategies for overcoming this limitation are elaborated in Section 3.2.

\section{Experimental Assessment}

The experimental setup utilized the following hardware and software specifications: an Intel Core i7 processor clocked at 2.3GHz, 32GB of RAM, operating on a Windows 10 system with Python 3. \\
The datasets for the performance metrics were randomly generated. The ratio of "1"s in the output column of the truth table, referred to as density, serves as a benchmark for the algorithm's combinatorial optimization capabilities. \\
Performance was evaluated across four varying density levels: 0.1, 0.01, 0.001, and 0.0001, and across different dimensions of input. The performance metrics were recorded in units of time, specifically seconds.

\begin{table}[H]
\resizebox{\linewidth}{!}{
\begin{tabular}{|l|llll|}
\hline
Dimensions\textbackslash{}Density & 0.1    & 0.2    & 0.3    & 0.4    \\ \hline
10                               & 0.0011 & 0.0028 & 0.0059 & 0.0135 \\
11                               & 0.0023 & 0.0063 & 0.0153 & 0.0353 \\
12                               & 0.0054 & 0.0155 & 0.0376 & 0.1057 \\
13                               & 0.0118 & 0.0348 & 0.0925 & 0.3115 \\
14                               & 0.0259 & 0.0788 & 0.2446 & 0.968  \\
15                               & 0.056  & 0.1778 & 0.6252 & 2.9035 \\ \hline
\end{tabular}
}
\caption{The density level is 0.1. Test results are measured in seconds.}
\label{LABEL 1}
\end{table}

\begin{table}[H]
\resizebox{\linewidth}{!}{
\begin{tabular}{|l|llll|}
\hline
Dimensions\textbackslash{}Density & 0.002   & 0.004   & 0.006   & 0.008   \\ \hline
14                               & 0.0010 & 0.0014 & 0.0020 & 0.0024 \\
15                               & 0.0024 & 0.0031 & 0.0041 & 0.0051 \\
16                               & 0.0050 & 0.0065 & 0.0084 & 0.0102 \\
17                               & 0.0099 & 0.0134 & 0.0172 & 0.0219 \\
18                               & 0.0200 & 0.0282 & 0.0354 & 0.0454 \\
19                               & 0.0406 & 0.0573 & 0.0747 & 0.0930 \\
20                               & 0.0836 & 0.1206 & 0.1531 & 0.1936 \\
21                               & 0.1687 & 0.2457 & 0.3259 & 0.3992 \\
22                               & 0.3458 & 0.5067 & 0.6613 & 0.8393 \\
23                               & 0.7054 & 1.0538 & 1.3900 & 1.7406 \\ \hline
\end{tabular}
}
\caption{The density level is 0.001. Test results are measured in seconds.}
\label{LABEL 3}
\end{table}

\begin{table}[H]
\centering
\resizebox{\columnwidth}{!}{%
\begin{tabular}{|l|llll|}
\hline
\multicolumn{1}{|c|}{Dimensions\textbackslash{}Density} & 0.0002   & 0.0004   & 0.0006   & 0.0008   \\ \hline
14                                          & 0.0008 & 0.0008 & 0.0009 & 0.0010 \\
15                                          & 0.0016 & 0.0014 & 0.0018 & 0.0019 \\
16                                          & 0.0033 & 0.0032 & 0.0039 & 0.0038 \\
17                                          & 0.0060 & 0.0065 & 0.0075 & 0.0076 \\
18                                          & 0.0122 & 0.0139 & 0.0152 & 0.0156 \\
19                                          & 0.0251 & 0.0268 & 0.0318 & 0.0314 \\
20                                          & 0.0488 & 0.0529 & 0.0634 & 0.0626 \\
21                                          & 0.0971 & 0.1044 & 0.1283 & 0.1301 \\
22                                          & 0.1977 & 0.2108 & 0.2571 & 0.2538 \\
23                                          & 0.3913 & 0.4657 & 0.5189 & 0.4898 \\ \hline
\end{tabular}
}
\caption{The density level is 0.0001. Test results are measured in seconds.}
\label{LABEL 4}
\end{table}

The subsequent experimentation utilizes the IWLS93 benchmark, employing the approach developed by Ammes and Lau Neto \cite{ANalgorithm} (henceforth referred to as AN's approach) as a benchmark for comparison. The results of this analysis are presented in Table 7. It is important to note that our algorithm is specifically designed for single-output formats. For these multi-output benchmarks, we implemented the following procedure: all output expressions were consolidated into a single representation, eliminating redundant implicates and removing those that could be decomposed into two sub-implicates, each of which is already present in the outputs. Regarding the data, due to memory constraints, AN's approach involved a selective modification of the dataset. In contrast, our methodology employed the original, unaltered dataset, leading to variations in the 'orig.' values presented in the table.\\
Our algorithm demonstrates significant improvements in terms of computation time and exhibits commendable performance across the majority of datasets. However, it is noteworthy that for some datasets, the reduction efficiency is comparatively lower. Unlike AN's approach, our method does not consider the Number of Errors (NoE) and is more memory-efficient. This makes it universally applicable to any input and output size without concerns of excessive errors or computational memory limitations.\\

\begin{table}[!ht]
\resizebox{\columnwidth}{!}{%
\begin{tabular}{ccccllcll}
\hline
\multicolumn{1}{|c|}{\multirow{3}{*}{Circuit}} & \multicolumn{8}{c|}{Literals}                                                                                                                                                                                                              \\ \cline{2-9} 
\multicolumn{1}{|c|}{}                         & \multicolumn{3}{c|}{AN's}                                                            & \multicolumn{3}{l|}{Ours}                                                           & \multicolumn{2}{l|}{Reduction \%}                             \\ \cline{2-9} 
\multicolumn{1}{|c|}{}                         & \multicolumn{1}{c|}{Orig.} & \multicolumn{1}{c|}{Num} & \multicolumn{1}{c|}{ER}      & \multicolumn{1}{l|}{Orig.} & \multicolumn{1}{c|}{Num} & \multicolumn{1}{c|}{ER}     & \multicolumn{1}{l|}{AN's} & \multicolumn{1}{l|}{Ours}         \\ \hline
\multicolumn{1}{|c|}{alu4(i:14;o:8)}           & 5087                       & 4847                     & \multicolumn{1}{c|}{0.09\%}  & 8903                       & 3913                     & \multicolumn{1}{c|}{0.76\%} & 4.717908394               & \multicolumn{1}{l|}{56.04852297}  \\
\multicolumn{1}{|c|}{apex4(i:9;o:19)}          & 5419                       & 5024                     & \multicolumn{1}{c|}{3.12\%}  & 5435                       & 6002                     & \multicolumn{1}{c|}{1.12\%} & 7.289167743               & \multicolumn{1}{l|}{-10.4323827}  \\
\multicolumn{1}{|c|}{b12(i:15;o:9)}            & 207                        & 207                      & \multicolumn{1}{c|}{0.04\%}  & 2303                       & 142                      & \multicolumn{1}{c|}{0.14\%} & 0                         & \multicolumn{1}{l|}{93.8341294}   \\
\multicolumn{1}{|c|}{clip(i:9;o:5)}            & 793                        & 584                      & \multicolumn{1}{c|}{3.12\%}  & 1055                       & 576                      & \multicolumn{1}{c|}{4.69\%} & 26.3556116                & \multicolumn{1}{l|}{45.4028436}   \\
\multicolumn{1}{|c|}{ex1010(i:10;o:10)}        & 2718                       & 2636                     & \multicolumn{1}{c|}{1.56\%}  & 11711                      & 2352                     & \multicolumn{1}{c|}{0.00\%} & 3.016924209               & \multicolumn{1}{l|}{79.91631799}  \\
\multicolumn{1}{|c|}{inc(i:7;o:9)}             & 198                        & 125                      & \multicolumn{1}{c|}{12.50\%} & 288                        & 147                      & \multicolumn{1}{c|}{0.00\%} & 36.86868687               & \multicolumn{1}{l|}{48.95833333}  \\
\multicolumn{1}{|c|}{misex3(i:14;o:14)}        & 7784                       & 7242                     & \multicolumn{1}{c|}{0.09\%}  & 19819                      & 4029                     & \multicolumn{1}{c|}{0.54\%} & 6.963001028               & \multicolumn{1}{l|}{79.67102276}  \\
\multicolumn{1}{|c|}{rd84(i:8;o:4)}            & 2070                       & 1511                     & \multicolumn{1}{c|}{6.25\%}  & 2459                       & 2323                     & \multicolumn{1}{c|}{0.00\%} & 27.00483092               & \multicolumn{1}{l|}{5.530703538}  \\
\multicolumn{1}{|c|}{sao2(i:10;o:4)}           & 496                        & 165                      & \multicolumn{1}{c|}{6.25\%}  & 501                        & 27                       & \multicolumn{1}{c|}{0.00\%} & 66.73387097               & \multicolumn{1}{l|}{94.61077844}  \\
\multicolumn{1}{|c|}{sqrt8(i:8;o:4)}           & 188                        & 83                       & \multicolumn{1}{c|}{6.25\%}  & 195                        & 148                      & \multicolumn{1}{c|}{1.27\%} & 55.85106383               & \multicolumn{1}{l|}{24.1025641}   \\
\multicolumn{1}{|c|}{table5(i:17;o:15)}        & 2501                       & 2270                     & \multicolumn{1}{c|}{0.01\%}  & 2502                       & 2989                     & \multicolumn{1}{c|}{0.00\%} & 9.236305478               & \multicolumn{1}{l|}{-19.46442846} \\ \hline
\multicolumn{1}{l}{}                           & \multicolumn{1}{l}{}       & \multicolumn{1}{l}{}     & \multicolumn{1}{l}{}         &                            &                          & \multicolumn{1}{l}{}        &                           &                                  
\end{tabular}%
}
\caption{}
\label{tab:my-table}
\end{table}

\begin{table}[!ht]
\resizebox{\columnwidth}{!}{%
\begin{tabular}{cllcc}
\hline
\multicolumn{1}{|c|}{\multirow{3}{*}{Circuit}} & \multicolumn{2}{l|}{\multirow{2}{*}{Literals Reduction \%}}   & \multicolumn{2}{c|}{\multirow{2}{*}{Time(s)}}             \\
\multicolumn{1}{|c|}{}                         & \multicolumn{2}{l|}{}                                         & \multicolumn{2}{c|}{}                                     \\ \cline{2-5} 
\multicolumn{1}{|c|}{}                         & \multicolumn{1}{l|}{AN's} & \multicolumn{1}{l|}{Ours}         & \multicolumn{1}{c|}{AN's} & \multicolumn{1}{c|}{ours}     \\ \hline
\multicolumn{1}{|c|}{alu4(i:14;o:8)}           & 4.717908394               & \multicolumn{1}{l|}{56.04852297}  & 9.73                      & \multicolumn{1}{c|}{0.500997} \\
\multicolumn{1}{|c|}{apex4(i:9;o:19)}          & 7.289167743               & \multicolumn{1}{l|}{-10.4323827}  & 22.08                     & \multicolumn{1}{c|}{0.039001} \\
\multicolumn{1}{|c|}{b12(i:15;o:9)}            & 0                         & \multicolumn{1}{l|}{93.8341294}   & 1.14                      & \multicolumn{1}{c|}{0.762967} \\
\multicolumn{1}{|c|}{clip(i:9;o:5)}            & 26.3556116                & \multicolumn{1}{l|}{45.4028436}   & 0.95                      & \multicolumn{1}{c|}{0.01}     \\
\multicolumn{1}{|c|}{ex1010(i:10;o:10)}        & 3.016924209               & \multicolumn{1}{l|}{79.91631799}  & 1.46                      & \multicolumn{1}{c|}{0.020999} \\
\multicolumn{1}{|c|}{inc(i:7;o:9)}             & 36.86868687               & \multicolumn{1}{l|}{48.95833333}  & 0.13                      & \multicolumn{1}{c|}{0.003999} \\
\multicolumn{1}{|c|}{misex3(i:14;o:14)}        & 6.963001028               & \multicolumn{1}{l|}{79.67102276}  & 8.08                      & \multicolumn{1}{c|}{5.877966} \\
\multicolumn{1}{|c|}{rd84(i:8;o:4)}            & 27.00483092               & \multicolumn{1}{l|}{5.530703538}  & 3.03                      & \multicolumn{1}{c|}{0.024033} \\
\multicolumn{1}{|c|}{sao2(i:10;o:4)}           & 66.73387097               & \multicolumn{1}{l|}{94.61077844}  & 1.04                      & \multicolumn{1}{c|}{0.022003} \\
\multicolumn{1}{|c|}{sqrt8(i:8;o:4)}           & 55.85106383               & \multicolumn{1}{l|}{24.1025641}   & 0.16                      & \multicolumn{1}{c|}{0.005969} \\
\multicolumn{1}{|c|}{table5(i:17;o:15)}        & 9.236305478               & \multicolumn{1}{l|}{-19.46442846} & 16.08                     & \multicolumn{1}{c|}{2.413}    \\ \hline
\multicolumn{1}{l}{}                           &                           &                                   & \multicolumn{1}{l}{}      & \multicolumn{1}{l}{}         
\end{tabular}%
}
\caption{}
\label{tab:my-table}
\end{table}

Furthermore, the performance of our algorithm was benchmarked against Song's algorithm \cite{songalgorithm}, using a random data set with a density of 0.1. The y-axis in the graph represents the time ratio as the number of inputs increases. The corresponding results are illustrated in Figure 1. It is evident from the analysis that our algorithm exhibits a lower time complexity, demonstrating enhanced performance with increasing numbers of inputs and outputs.\\

\begin{figure}[!ht]
  \centering
  \includegraphics[width=\columnwidth]{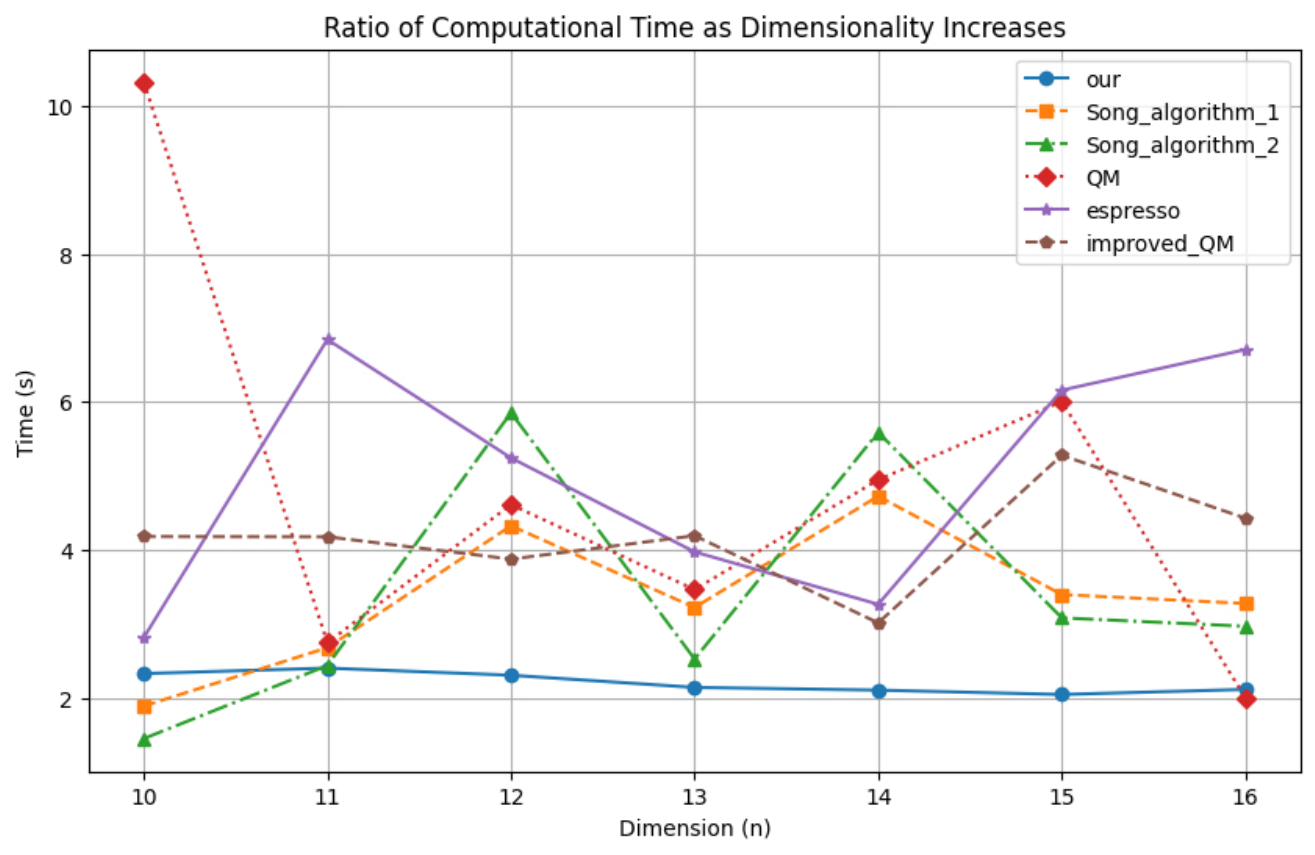}
  \caption{Rate of different algorithm}
\end{figure}

\section{Conclusion}
This study has unveiled the intrinsic interrelations within Boolean functions, particularly between implicates and elements, along with the associations among the elements themselves. While our algorithm was initially designed for single-output scenarios, it demonstrates commendable performance in multi-output situations as well, as detailed in the comparison with AN's approach. Notably, this method not only excels in speed, making it suitable for subsequent Boolean term reduction, but it can also be directly applied to truth tables at the design's inception for initial optimization. The Conditional Flood Fill algorithm presented in this article significantly simplifies Boolean functions in a non-approximate manner, ensuring high accuracy and rapid execution. The elapsed time ratio of approximately 2 indicates a low complexity, enabling effective performance even in high-dimension scenarios.

\end{document}